\documentstyle[sprocl,epsfig,here]{article}

\def\beq{\begin{equation}}
\def\eeq{\end{equation}}
\def\bea{\begin{eqnarray}}
\def\eea{\end{eqnarray}}
\def\bq{\begin{quote}}
\def\eq{\end{quote}}

\def\ASAS{{\it Astron. and Astrophys.} }

\def\NC{{\it Nuovo Cimento} }
\def\NP{{\it Nucl.Phys.} }
\def\PL{{\it Phys.Lett.} }
\def\PR{{\it Phys.Rev.} }
\def\PRL{{\it Phys.Rev.Lett.} }

\def\YF{{\it Yadernaya Fizika} }

\def\gappeq{\mathrel{\rlap {\raise.5ex\hbox{$>$}}
{\lower.5ex\hbox{$\sim$}}}}

\def\lappeq{\mathrel{\rlap{\raise.5ex\hbox{$<$}}
{\lower.5ex\hbox{$\sim$}}}}
\begin{document}
\pagestyle{empty}
\begin{flushright}
CERN-TH/96-325\\
hep-ph/9612209
\end{flushright}
\vspace*{5mm}
\begin{center}
{\bf OUTLOOK ON NEUTRINO PHYSICS$^*$} \\
\vspace*{0.5cm} 
{\bf John Ellis} \\
TH Division, CERN, 1211 Geneva 23, Switzerland \\
\vspace*{0.5cm}  
{\bf ABSTRACT} \\ 
\end{center}
\vspace*{5mm}
\noindent
Some of the hot topics in neutrino physics are discussed,
with particular emphasis on neutrino oscillations. After
proposing credibility criteria for assessing various
claimed effects, particular stress is laid on the solar
neutrino deficit, which seems unlikely to have an
astrophysical explanation. Comments are also made on the
possibility of atmospheric neutrino oscillations and on
the LSND experiment, as well as cosmological aspects of
neutrinos and neutralinos. Several of the central issues
in neutrino physics may be resolved by the new generation
of experiments now underway, such as CHORUS, NOMAD and Superkamiokande,
and in preparation, such as SNO and a new round of accelerator-
and reactor-based neutrino-oscillation experiments. At the end,
there is a brief review of ways in which 
present and future CERN experiments
may be able to contribute to answering outstanding
questions in neutrino physics.

\vspace*{0.3cm}
\noindent
\rule[.1in]{12cm}{.002in}

\noindent
${}^*$
Invited Talk presented at the Conclusion of the Neutrino 96 Conference, 
Helsinki, June 1996.
\vspace*{0.3cm}

\begin{flushleft} 
CERN-TH/96-325 \\
November 1996
\end{flushleft}
\vfill\eject

\setcounter{page}{1}
\pagestyle{plain}

\title{OUTLOOK ON NEUTRINO PHYSICS}

\author{John Ellis}

\address{TH Division, CERN, CH-1211 Geneva 23, Switzerland}

\maketitle\abstract{
Some of the hot topics in neutrino physics are discussed, with
particular emphasis on neutrino oscillations. After proposing credibility
criteria for assessing various claimed effects, particular stress is laid
on the solar neutrino deficit, which seems unlikely to have an
astrophysical explanation. Comments are also made on the
possibility of atmospheric neutrino oscillations and the LSND
experiment, as well as cosmological aspects of neutrinos and neutralinos.
several of the central issues in neutrino physics may be resolved by
the new generation of experiments now underway, such as CHORUS, NOMAD
and Superkamiokande, and in preparation, such as SNO and a new round of
accelerator- and reactor-based neutrino-oscillation experiments. At the
end, there is a brief review of ways in which present and future CERN
experiments may be able to contribute to answering outstanding questions
in neutrino physics.}

\section{Preamble}

The story so far is that we know there are three (and only three)
light neutrino species~\cite{LEP}: apart from that, all we have are upper 
limits on neutrino properties~\cite{PDG}. According to the Standard Model,
each neutrino flavour carries its own lepton number, and these are all
separately conserved: $\Delta L_e = \Delta L_{\mu} = \Delta L_{\tau} = 0$,
and all the light neutrinos are in fact massless:
$m_{\nu_e} = m_{\nu_{\mu}} = m_{\nu_{\tau}} = 0$. One may either
accept the Standard Model's opinion, and use neutrinos as tools for
probing nuclear/particle physics, astrophysics and cosmology, or
one may question the Standard Model's wisdom, and study neutrinos in their
own rights as possible arenas for new physics beyond the Standard Model.
Most of this meeting has been, and most of this talk is, concerned
with searches for such new physics, which is also the likely focus
of most future neutrino experiments, but let us first review briefly 
some Standard Model physics with neutrinos.

\section{Testing the Standard Model}

Neutrinos may be used in nuclear physics to measure interesting
matrix elements, which may cast light on nuclear wave functions.
In this connection, it was reported here~\cite{discrep} that the LSND
experiment
finds a discrepancy with calculations of $\sigma(\nu + {}^{12}C ->
\mu + N^*)$. Also, it should not be forgotten that neutrino-proton elastic
scattering may also cast
light on the presence of hidden strangeness in the proton~\cite{nup}.
It would be
a pity if these mundane issues got forgotten completely in the
enthusiasm for possible new physics with LSND.

Neutrinos can also be used to make 
measurements of fundamental Standard Model parameters, such as
sin$^2 \theta_W$ and $\alpha_s$. We heard here~\cite{CCFR} that the
present 
error on the deep-inelastic $\nu$-N scattering
measurement of sin$^2 \theta_W$ is $\pm 0.0048$, to be compared
with the LEP/SLC error of about $\pm 0.0003$. However, this
comparison is not really fair, since different definitions
of sin$^2 \theta_W$ are used. The $\nu$-N measurement can be
considered as effectively a determination of $M_W/M_Z$. The
promised~\cite{CCFR} error of $\pm 0.0025$ is equivalent to $\delta M_W =
\pm 130$ GeV, comparable to the present error on the direct
measurements of $M_W$~\cite{CDFD0,LEP2}, and should be compared with the
errors of $\pm 50$ MeV or so expected in the future
from LEP 2 and the Tevatron collider. These measurements 
are in turn important for the indirect determination of the
Higgs mass. A global fit to all the available electroweak
measurements, including present $\nu$-N scattering data, gives~\cite{MH}
\begin{equation}
\hbox{log}_{10} (M_H/\hbox{GeV}) \, = \, 2.16 \pm 0.33
\label{mhiggs}
\end{equation}
corresponding to a factor of 2 error in $M_H$, which we may hope
future $\nu$-N scattering experiments could help reduce.

Deep-inelastic $\nu$-N scattering also provides an
interesting determination of $\alpha_s$, in particular from the 
Gross-Llewellyn-Smith sum rule~\cite{GLS}:
\begin{equation}
\alpha_s(M_Z) \, = \, 0.111 \pm 0.003 \pm 0.004
\label{alphas}
\end{equation}
where the dominant experimental systematic error comes
from the beam energy calibration~\cite{CCFR}. Taking this into account,
the measurement (\ref{alphas}) is not incompatible with the world
average of $\alpha_s(M_Z) \simeq 0.118 \pm 0.003$~\cite{Schmelling}. There
are
other low-energy determinations of $\alpha_s(M_Z)$
which yield relatively high values, e.g., from $\tau$ decay~\cite{tau} and
the
Bjorken sum rule~\cite{EGKS}, and the available measurements
seem to me to scatter normally around this central value, with no
significant trend for lower-energy determinations to yield lower values,
as has sometimes been suggested. To make a graceful transition to
physics beyond the Standard Model, it is worth recalling that
measurements of $\alpha_s$ and sin$^2
\theta_W$ test theories of Grand Unification, so these
neutrino experiments bear on fundamental theoretical issues.

\section{Massive Neutrinos?}

As already mentioned, the Standard Model abhors masses for
the neutrinos, and all we have so far from experiments
are upper limits~\cite{PDG}:
\begin{equation}
m_{\nu_e}\,<\,4.5 \hbox{eV} \, ?,\,m_{\nu_{\mu}}\,<\,160 \hbox{KeV},
\,m_{\nu_{\tau}}\,<\,23 \hbox{MeV}
\end{equation}
Theoretically, there is no good reason for the neutrino masses to vanish,
and non-zero masses are expected in generic grand unification theories.
There are probably more models for the neutrino mass matrix that there are
theorists who have worked on the problem: here I shall be guided by the
simplest see-saw form~\cite{seesaw}:
\begin{equation}
(\nu_L, \bar\nu_R)~~\left(\matrix{\sim 0 & \sim m_q \cr \sim m_q & \sim
M_{GUT}}\right)~~\left(\matrix{\nu_L\cr\bar\nu_R}\right)
\label{seesaw}
\end{equation}
whose diagonalization leads naturally to light neutrinos
\begin{equation}
m_{\nu_i}\sim {m_{q_i}^2 \over m_{GUT}} << m_{q,\ell}
\label{smallmass}
\end{equation}
These non-zero masses would be accompanied by mixing via angles
$\theta_{ij}: i,j = e, \mu, \tau$ analogous (and in some models related)
to the Cabibbo-Kobayashi-Maskawa angles in the quark sector.

Our best direct information about $m_{\nu_e}$ comes from measurements of
the end point of the ${}^3$He $\beta$-decay spectrum.  The latest results
of two current experiments were presented here.  As seen in Fig.~1,
neither data set is fit well by the standard theory.  Relative to the
standard Kurie plot, the Mainz spectrum~\cite{Bonn} shown in Fig.~1(a)
has a banana shape that raises question about the presence of
final-state ionic and/or molecular effects that are not well understood.
In the case of the Troitsk experiment~\cite{Lobashev} (see Fig.~1(b)),
there is a
mysterious bump near the end of the spectrum that appears at different
energies in the data from different years.  Both experiments 
report~\cite{Bonn,Lobashev} a feature around $250$ to $300$ eV from the end
point, with an apparent branching probability of $1$ or $2\%$.  However,
before accepting a very stringent upper limit on $m_{\nu_e}$ 
from the end-point measurements, or crying
$\nu_h$ on the basis of this lower-energy feature, the community will
require more evidence that the theoretical
understanding of the spectrum is adequate.

The Mainz group does plan~\cite{Bonn} to measure final-state effects by
monitoring additional $\gamma$s and molecular break up in a more
calrimetric approach. There is also an
interesting suggestion~\cite{Swift} to use ${}^{187}$Re (which has $Q =
2$ KeV) in a calorimetric experiment based on superconducting bolometry to
remove final-state effects.

A powerful indirect way to search for $m_{\nu_e}$ is to look for $0\nu$
$\beta\beta$ decay.  However, the interpretation of these experiments is
subject to uncertainties: a positive effect could be due to some other
mechanism such as $R$ violation in supersymmetric models, and, even if the
results continue to be negative, there could be a Dirac mass or some other
mechanism for a cancellation in $<m_{\nu_e}>$. Nevertheless, the
upper limit~\cite{KKGH}
\begin{equation}
<m_{\nu_e}> \lappeq 0.5 \, \hbox{eV}
\label{twobeta}
\end{equation}
is very impressive.  As we heard here~\cite{betabeta}, there are prospects
for incremental
improvements, leading perhaps to $m_{\nu_e} \lappeq 0.2 (0.1)$ eV by the
year 2000 (2005), and many other horses will be competing in this race.

\begin{figure}[H]
\hglue2cm
 \epsfig{figure=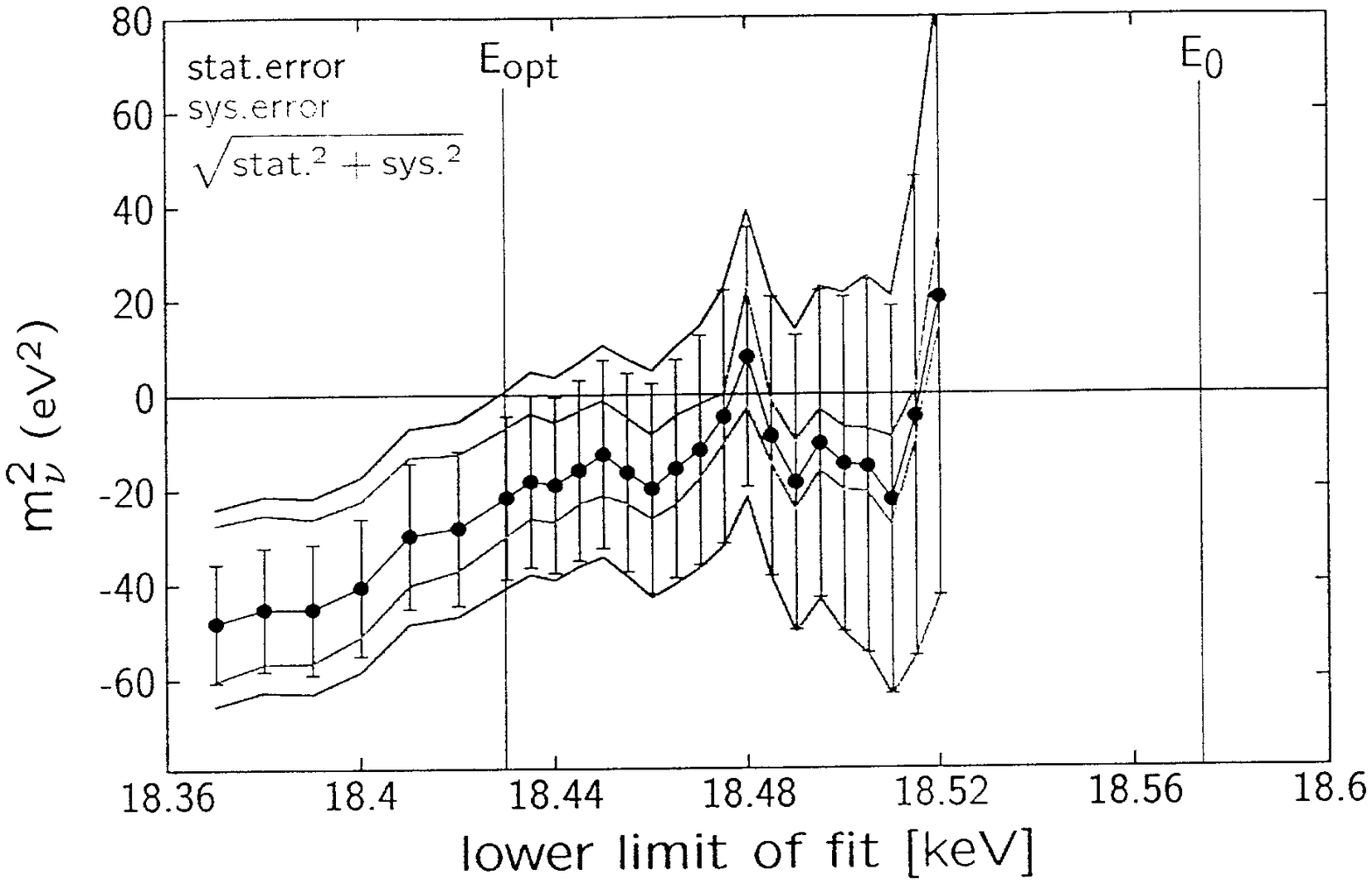,width=7cm}
\vglue.5cm
\hglue2cm
 \epsfig{figure=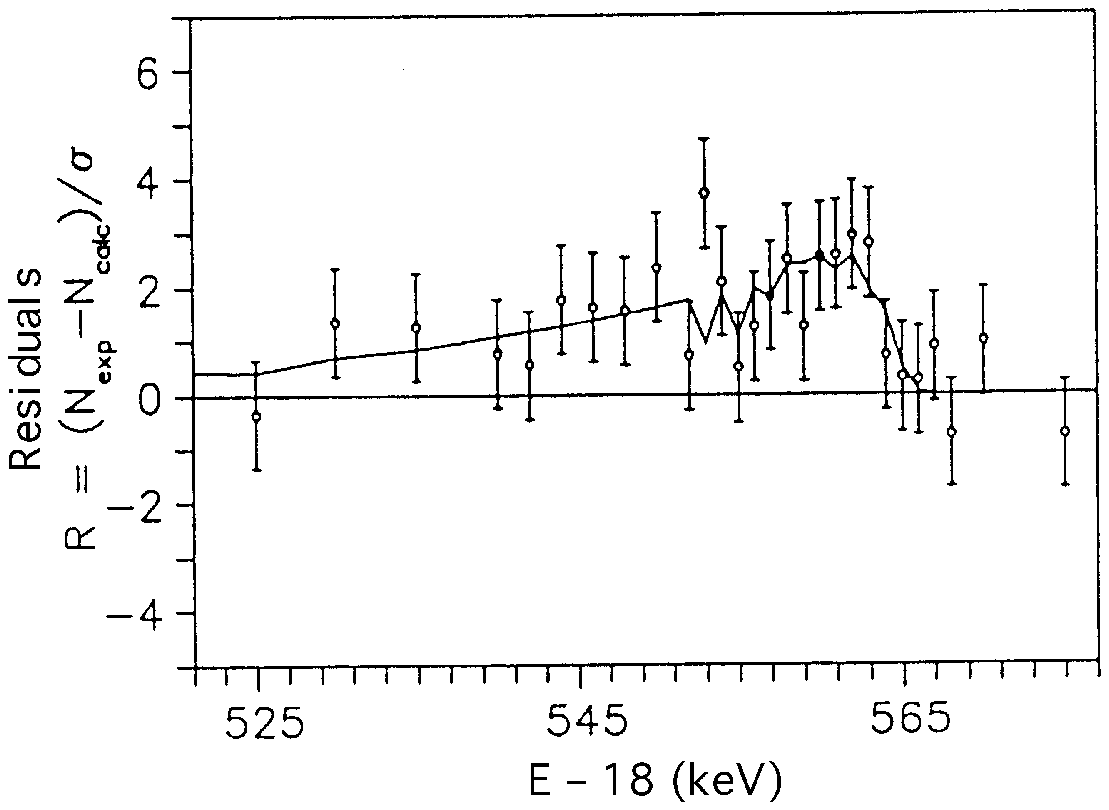,width=7cm}
\caption[]{Features at the ends of the tritium $\beta$-decay spectra seen (a) in
the Mainz experiment~\cite{Bonn}, and (b) in the Troitsk
experiment~\cite{Lobashev}, which are not fitted well by standard models.}
\end{figure}

\newpage
\section{Neutrino Oscillations?}

This is a very difficult experimental field, and the stakes are very high.
Therefore, we must ask for a very high standard of proof: my personal
credibility criteria are that any claimed effect should be confirmed by
more than one experiment using more than one technique.  From this
standpoint, the possibility of {\bf Solar Neutrino Oscillations} holds up
very well: the solar neutrino deficit has now been seen by $5$ experiments
(Homestake~\cite{Home},
Kamiokande~\cite{Suzuki}, SAGE~\cite{Gavrin}, GALLEX~\cite{Kirsten} and
now Superkamiokande~\cite{Suzuki}) using $3~1/2$ techniques (Cl, H$_2$O
and two somewhat different Ge techniques).  The claim of {\bf Atmospheric
Neutrino Oscillations}, on the other hand, is supported by just $2~1/2$
experiments (Kamiokande~\cite{Suzuki} and IMB~\cite{IMB}, with
Soudan~\cite{Soudan} as yet inconclusive),
using just $1~1/2$ techniques (water Cerenkov
and $1/2$ for calorimetry/tracking, since other tracking detectors (NUSEX,
Fr\'ejus) do not
see the effect).  Finally, the claim of possible {\bf Accelerator
Neutrino Oscillations} is made by only one experiment
(LSND~\cite{Caldwell}) using a scintillator technique.

In my view, the fact of a solar neutrino deficit is well
established~\cite{Bahcall}, and
the essential question now is whether it is due to astrophysics or
neutrino properties.  On the other hand, the existence of an atmospheric
neutrino deficit cannot be regarded as well established, but requires
further checks.  There are still considerable uncertainties in the
absolute cosmic-ray neutrino fluxes~\cite{Gaisser} which cloud the
interpretation (too few $\nu_{\mu}$ or too many $\nu_e$?), and
confirmation
by different techniques is essential.  The LSND~\cite{Caldwell} claim of
an
accelerator neutrino effect is in even greater need of confirmation, and
it is encouraging that the LSND Collaboration itself, as well as
an upgrade of the KARMEN experiment~\cite{KARMEN}, should be able
to address this issue.

We saw at this meeting new, and in some cases definitive, results from the
four experiments that see a solar neutrino deficit.  Homestake now
reports~\cite{Home} a flux of $2.54\pm 0.16\pm 0.14$ SNU, and the previous
suggestion of some time dependence (an anti-correlation with the sunspot
number) has declined substantially in statistical significance.  The
definitive Kamiokande result of $0.424\pm 0.029\pm 0.05$ of the
Bahcall-Pinsonneault 1995 flux also exhibits no significant time
dependence~\cite{Suzuki}:
\begin{equation}
{Data\over Standard~Solar~Model}= 0.398^{+0.088}_{-0.078} +
(9.4^{+7.2}_{-7.0}) 10^{-4}\times N_{SS}
\label{sunspots}
\end{equation}
SAGE~\cite{Gavrin} reported here a new result:
$72^{+15 +5}_{-10 -7}$ SNU and the preliminary results of a calibration
test:
$\epsilon = 0.54\pm 0.11$.  We also heard about new data from
GALLEX~\cite{Kirsten}, which are somewhat below their previous average,
though consistent within the errors.  Their latest result is $70 \pm 8$
SNU, which includes statistical and systematic errors, the latter
incorporating the results from their two calibration runs, which together
yield $\epsilon = 0.92\pm 0.07$~\cite{Kirsten}. 
These two new results tend even to sharpen
the dilemma of the solar neutrino deficit.

As was discussed here by Smirnov~\cite{Smirnov}, 
and seen in Fig.~2, the indications are that
the deficit is not monotonic with neutrino energy, in the way that would
be expected, for instance, if it was due simply to a reduction in the
central temperature of the Sun.  Nor does the deficit look
energy-independent~\cite{Fior}, as was suggested here by
Conforto~\cite{Conforto}.
The latest Standard Solar Model calculations do include diffusion effects
and the latest available opacities~\cite{Bahcall}. Personally, I am at
least reassured
about the accuracy of the assumed values of the nuclear interaction rates,
but it still seems very difficult to accommodate the apparent (near?)
absence of intermediate-energy solar neutrinos coming mainly from
${}^7$Be.

Since there was no talk here on helioseismology, I would like to underline
that this does not help solve the solar neutrino problem~\cite{Dar}, but
rather the
reverse~\cite{Betal}.  Quoting from~\cite{helio}: ``substantial mixing...
solution of...
deficit of ${}^{37}$C1... seems unlikely", ``inclusion of settling...
tends
to increase neutrino fluxes", ``helioseismological results accentuate the
neutrino problem", ``models that... reduce the neutrino flux... are
generally inconsistent with the observed frequencies", and ``it appears
unlikely that the solar neutrino problem will find an astrophysical
solution".

\begin{figure}[H]
\hglue2cm
 \epsfig{figure=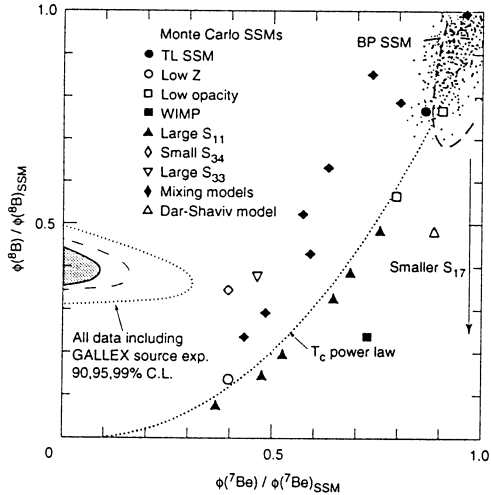,width=7cm}
\caption[]{A planar presentation~\cite{Fior} of the solar neutrino deficits
seen in different experiments, compared with a selection of different
models. The data do not support a suppression that is independent of
energy~\cite{Conforto}.}
\end{figure}

As is well known, and was emphasized here by Smirnov~\cite{Smirnov} and by
Petcov~\cite{Petcov}, matter-enhanced
Mikheyev-Smirnov-Wolfenstein (MSW)~\cite{MSW} oscillations
fit the data very well: a fit~\cite{Krastev} 
including the latest data is
shown in Fig.~3(a).  An issue raised here was the possibility of a
diminution of the MSW effect by fluctuations in the solar
density~\cite{Burgess,Rossi}: it will be important to take
helioseismological constraints into account when evaluating this
possibility.  We should also
not forget the possibility of vacuum oscillations:  the results of a
fit~\cite{Krastev} including all the latest data are shown in Fig.~3(b).
Perhaps our biggest peril is theoretical seduction by the MSW~\cite{MSW}
idea: {\it caveat emptor}, and do not forget that ``the Sun is not a piece
of cake"~\cite{Rowley}!

\begin{figure}[H]
\mbox{\epsfig{figure=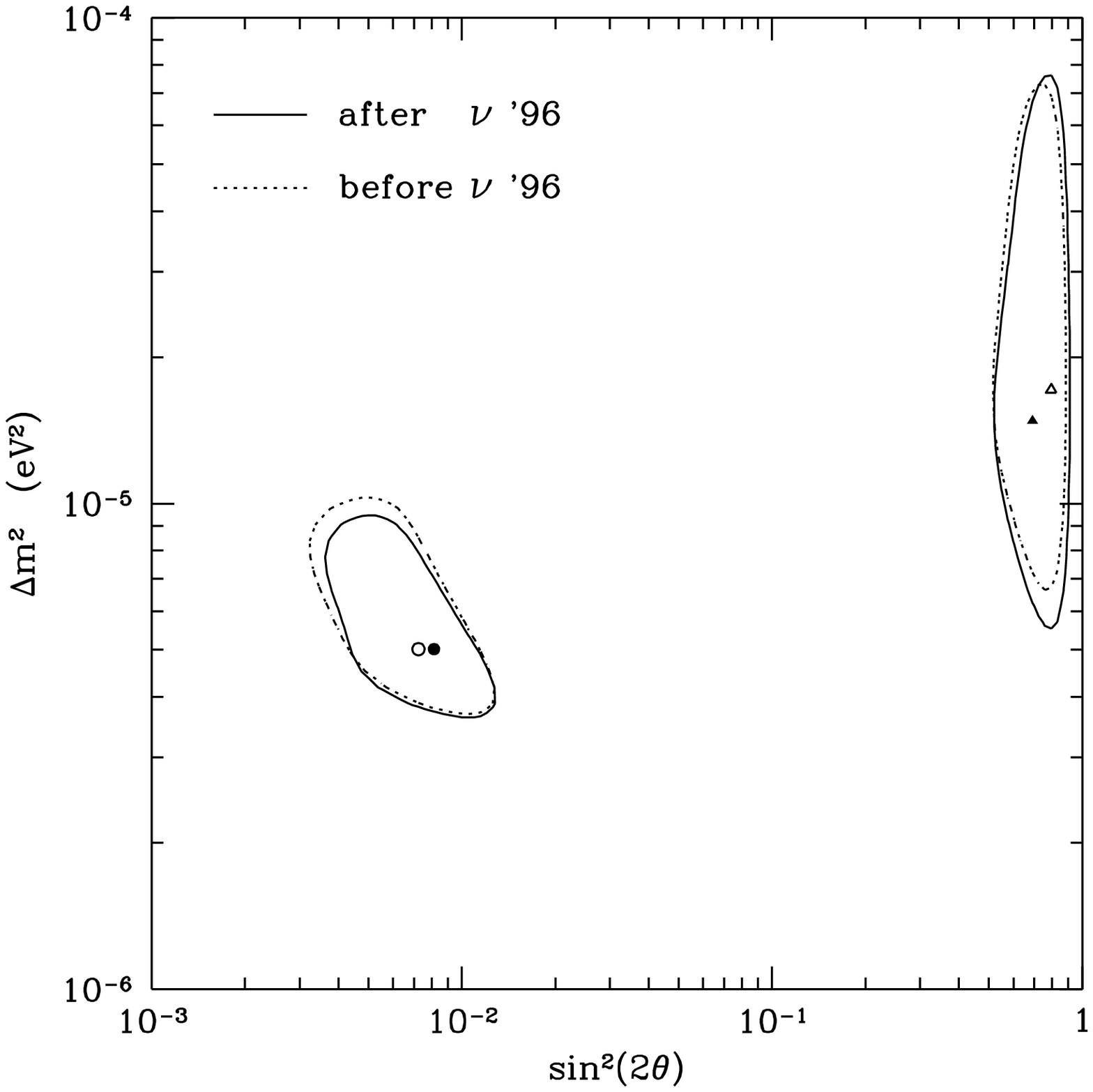,width=5.5cm}(a)}
\mbox{\epsfig{figure=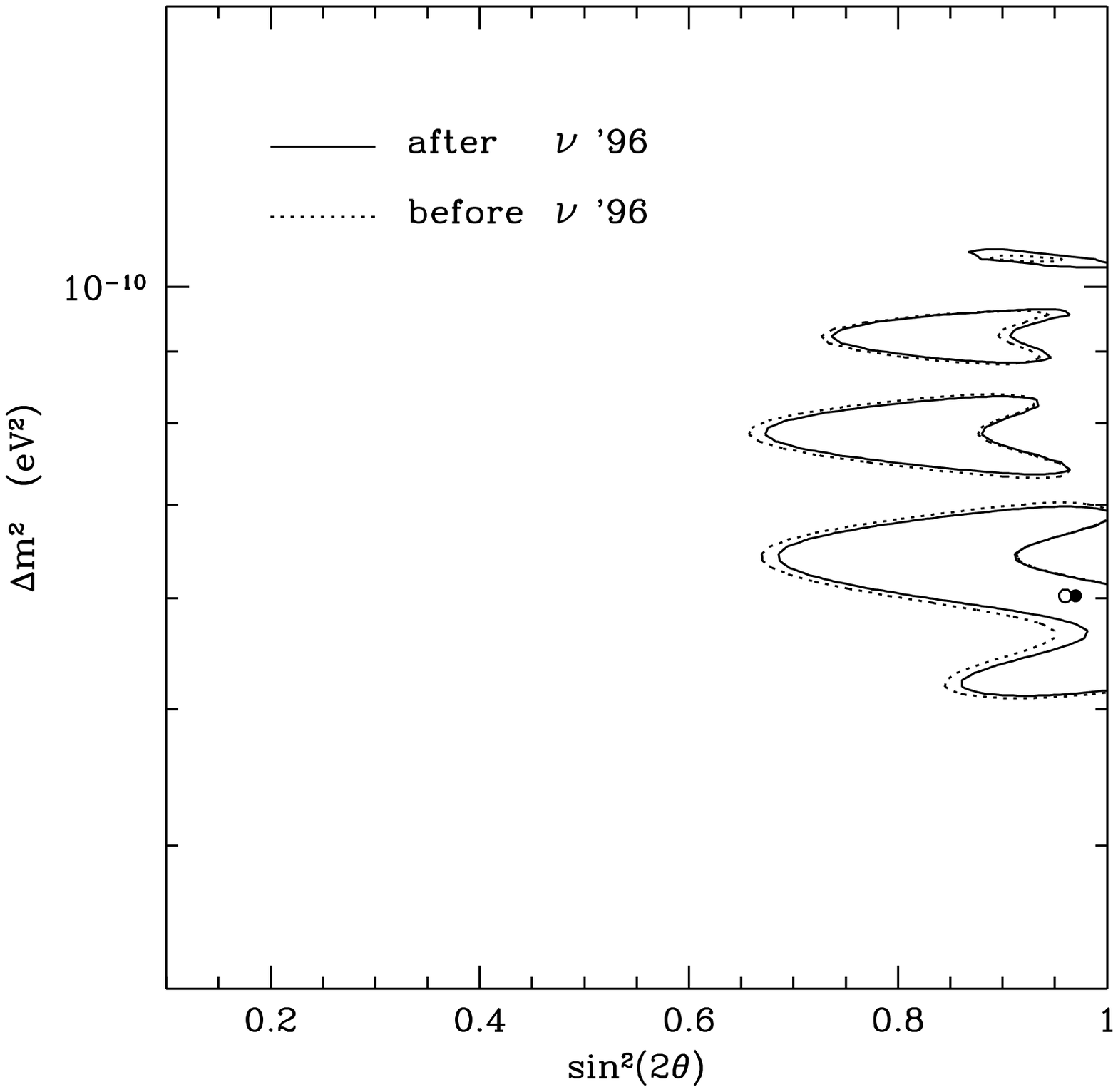,width=5.5cm}(b)}
\caption[]{Fits~\cite{Krastev} to the solar neutrino data available before and
after this meeting: (a) within an MSW interpretation, and (b) using vacuum
oscillations.}
\end{figure}

The outlook for progress on the solar neutrino problem is good.  The first
experiment in the next generation, Superkamiokande, is already operating,
and its preliminary data confirm the previous measurements of the
Kamiokande experiment~\cite{Suzuki}.  It should soon supplement this with
additional information on the ${}^8$B neutrino spectrum, and possible
seasonal and day/night effects, thereby providing us with many tests of
the MSW interpretation of the solar neutrino deficit.  The Homestake
Iodine experiment is about to start taking data~\cite{Home}.  During 1997,
SNO should enter into operation~\cite{SNO}, and provide us with the first
information on the neutral/charged current ratio, as well as more
detailed information on the possible spectrum distortion and the
possible seasonal effect, providing many tests of the
MSW and vacuum oscillation interpretations.  Also approved is the BOREXINO
experiment~\cite{BOREXINO}, which should provide us with definitive
information on the
intermediate-energy (${}^7$Be) neutrinos, and hence on the MSW
interpretation, starting in 1999.  Other projects for the Gran Sasso
laboratory are under active discussion.  One is the Gallium Neutrino
Observatory (GNO)~\cite{Kirsten}, which would continue the work of GALLEX,
eventually
increasing the detector mass to $100$ t.  The ICARUS
experiment~\cite{ICARUS} has
already been approved for a prototype module, and should eventually be
able to contribute a measurement of the ${}^8$B $\nu$ spectrum.  The
HELLAZ project~\cite{HELLAZ} would be able to contribute a measurement of
the $pp$ $\nu$
spectrum and test the vacuum oscillation interpretation.  It may take a
few years, but we are on the way to an experimental resolution of the
solar neutrino problem.

As already mentioned, a deficit in the atmospheric $\nu_{\mu} / \nu_e$
ratio has been reported by the H$_2$O Cerenkov detector Kamiokande:
\begin{equation}
{(\mu / e)_{data} \over (\mu / e)_{Monte~Carlo}} = 0.60^{+0.06}_{-0.05}
\label{Kamratio}
\end{equation}
in their sub-GeV data, and an angle-dependent effect has been seen in
their multi-GeV data~\cite{Suzuki}. A deficit in this ratio has also been
seen by the other H$_2$O detector IMB~\cite{IMB}, but is not
confirmed by the electronic detectors NUSEX
and Fr\'ejus~\cite{others}, whilst the Soudan result~\cite{Soudan}
\begin{equation}
{(\mu / e)_{data} \over (\mu / e)_{Monte~Carlo}} = 0.72 \pm
0.19^{+0.05}_{-0.07} 
\label{Soudratio}
\end{equation}
is ambiguous.  Because of uncertainties of as much as $30\%$ in the
absolute neutrino flux normalizations~\cite{Gaisser}, 
as seen in Fig.~4, the correct
interpretation of the ratios (\ref{Kamratio}, \ref{Soudratio}) is not
immediately obvious: are there too few $\nu_{\mu}$, or too many $\nu_e$,
or both? The ambiguity could be removed by more accurate data
on $\pi^{\pm}$ production in laboratory $p$-Nucleus collisions, which
should soon be available from the SPY experiment at CERN~\cite{SPY}, and
on the
cosmic-ray muon flux, which could be obtained with the L3 experiment at
CERN~\cite{cosmoL3}.

There are several prospects for experimental progress on the atmospheric
neutrino problem.  Superkamiokande~\cite{Suzuki} will soon have an order
of magnitude
more data than Kamiokande, with a correspondingly better control of
systematics, less leakage from the side, more stopping muons, and better
multi-GeV information.  However, it does not use a new technique.  There
are also several prospects of checks using accelerator neutrino beams over
long
base lines.  First among these will be the LBLE experiment~\cite{Suzuki} 
in which a
neutrino beam is sent from KEK to Superkamiokande, a distance of $250$km,
produced initially by a $12$ GeV $p$ beam in $1999$, which can look
directly for $\nu_{\mu} \rightarrow \nu_e$ oscillations, and subsequently
with a $50$ GeV $p$ beam in $2003$, which will permit the observation of
$\nu_{\tau}$.  In the United States, the MINOS
experiment~\cite{MINOS}
involves sending the $\nu$ beam produced by a $120$ GeV $p$ beam over
$730$ km to Soudan, which is currently scheduled to start in $2002$.  This
experiment should be sensitive to $\Delta m^2 = 0.001$ eV$^2$ and sin$^2
2\theta= 0.02$.  Also under discussion is a possible CERN-Gran Sasso
experiment~\cite{long}, which I will discuss later.

\begin{figure}[H]
\hglue2cm
\epsfig{figure=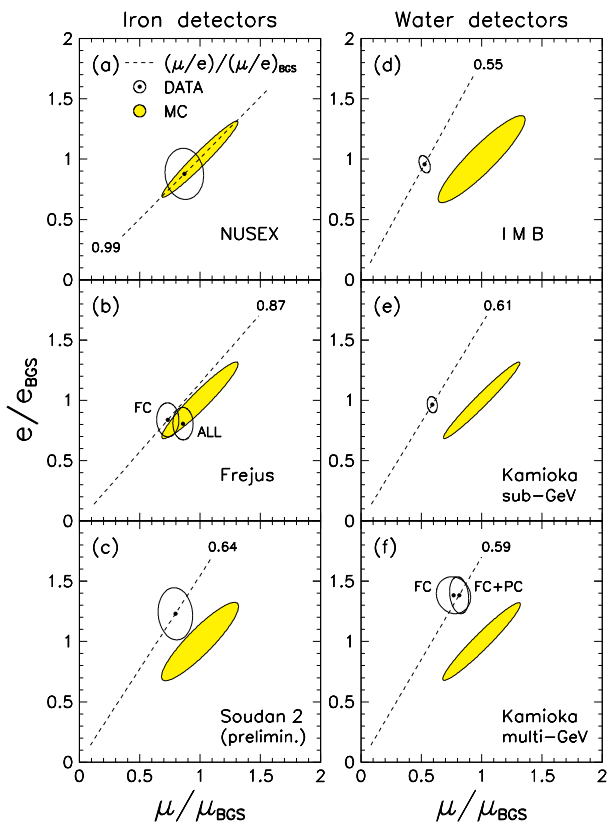,width=8cm}
\caption[]{Comparison~\cite{Fogli} of the atmospheric $\nu_{\mu}$ and $\nu_e$
fluxes observed in different experiments with theoretical
calculations~\cite{Gaisser}.}
\end{figure}

Some aspects of the atmospheric neutrino problem can be checked with
reactor experiments that are sensitive to $\nu_e$ disappearance.  The
Chooz experiment~\cite{Chooz} with a base line of $1$ km starts taking
data in $1996$,
and the Palo Verde experiment~\cite{Palo} with a $750$-m base line in
$1997$: they
should each reach $\Delta m^2 = 0.001$ and sin$^2 2\theta = 0.05$.

One of the most dramatic recent claims in neutrino physis has been that by
the LSND collaboration~\cite{Caldwell} to have observed an excess of
${\bar \nu_e}p \rightarrow e^+ n$ events (tagged by a second signature
from
$n p \rightarrow d \gamma$) which they interpret as evidence for ${\bar
\nu}_{\mu} \rightarrow {\bar \nu}_e$ oscillations.  The largest
irreducible
background is that due to a contamination of $8 \times 10^{-4}$ ${\bar
\nu}_e$ in the beam.  There is in addition a reducible background due to
cosmic rays, which is fought by shielding, veto counters, etc.  The LSND
collaboration reports a raw $e^+$ excess for $36$ MeV $< E < 60$ MeV
(recall that the ${\bar \nu}_{\mu}$ end point is at $53$ MeV) of $300 -
160.5 - 76.2 = 63.3 \pm 20$ events (where the first subtraction is for the
beam-off background, the second for the beam-on background), a $3-\sigma$
effect.  When they further select events with a $\gamma$ coincidence, as
determined by a cut on a function $R(\Delta t, N_{PMT}, \Delta r) > 30$,
as seen in Fig.~5,
they are left with $22 - (2.5 \pm 0.4) - (2.1 \pm 0.4) = 17.6 \pm 4.7$
events, somewhat more than $3\sigma$.

\begin{figure}[H]
\hglue2cm
\epsfig{figure=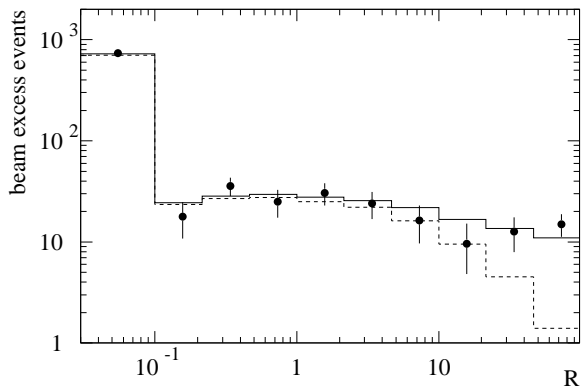,width=8cm}
\caption[]{The  tail of the $R$ distribution for events with $20
\hbox{MeV} < E_e < 60$ MeV in the LSND experiment has an apparent excess~\cite{Caldwell}.}
\end{figure}

There were initially some worries about the possibility of neutrons
leaking into the outer regions of the detector~\cite{Hill}.  If one
restricts the fiducial volume to the ``safest" $45~\%$, the signal is
reduced to $6$ events with a background of $1.7\pm 0.3$~\cite{Caldwell}.
It should also be noted~\cite{Caldwell} that the excess in $R$ shows up
only in the last two bins shown in Fig.~5.  Finally, the LSND result is
only marginally
consistent with previous experiments, most importantly $E776$,
KARMEN~\cite{KARMEN} and the Bugey reactor experiment, 
and the most recent CCFR limit~\cite{CCFRosc}, as seen in Fig.~6.

The LSND speaker here~\cite{Caldwell} talked of a "significant
oscillation-like excess that needs confirmation".  Many people here would
echo heartily the latter phrase.  Fortunately, efforts to confirm the
claimed effect are underway by LSND itself~\cite{Caldwell}, which is
searching in its data
for $\nu_{\mu} \rightarrow \nu_e$ oscillations, and should have results by
1997, and by the KARMEN collaboration~\cite{KARMEN}, which is improving
its $n$
detection efficiency and installing veto counters in its surrounding steel
blockhouse to veto cosmic-ray muons that might produce neutrons, and
should have results by 1998~\footnote{The KARMEN collaboration
presented~\cite{KARMEN} their own ``anomaly", namely an apparent excess of
$112\pm 32$ $\gamma$ events occurring after $3.6 \pm 0.25$ $\mu$s, that
might be due to the radiative decay of some state $X$ with a mass of
$33.9$ MeV.  However, this effect has not been confirmed by an experiment
at PSI~\cite{PSI}.}.

\begin{figure}[H]
\hglue2cm
\epsfig{figure=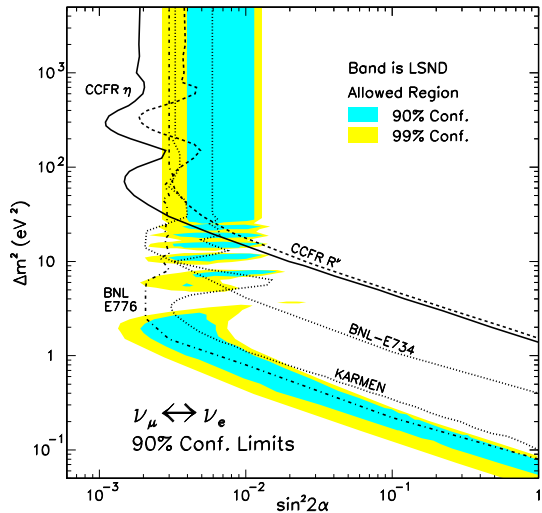,width=8cm}
\caption[]{Comparison of the LSND result~\cite{Caldwell} with those of $E776$,
KARMEN~\cite{KARMEN}, Bugey and CCFR~\cite{CCFRosc}.}
\end{figure}

Finally, let us recall the motivation for a new round of accelerator
neutrino-oscillation experiments.  Theorists of cosmological structure
formation would like some Hot Dark Matter~\cite{Primack}, which is most
plausibly one or
more massive neutrinos with $m_{\nu} \sim 1$ to $10$ eV, which is
compatible with the simplest possible seesaw model:
\begin{equation}
m_{\nu_{\tau}} \sim ({m_t \over m_c})^2 \times m_{\nu_{\mu}}
\label{munutau}
\end{equation}
It is, furthermore, plausible that the $\nu_{\tau}$ might have observable
mixing with the $\nu_{\mu}$.  These considerations
have motivated the CHORUS~\cite{CHORUS} and NOMAD~\cite{NOMAD} 
experiments, which will be running at
CERN until the end of 1997, and should attain sensitivities $\sim 0.0003$
to sin$^2\theta_{\mu\tau}$ for $\delta m^2 \sim 50$ eV$^2$. 
In my view, there is certainly
interest in extending this search down to lower $\Delta m^2$, as well as
to smaller mixing angles, by using either a longer baseline or a
lower-energy beam. Indeed, in the longer
term, the COSMOS experiment~\cite{COSMOS}
plans to take data for $4$ years starting in 2001, and expects to reach a
sensitivity to sin$^2\theta_{\mu\tau} \sim 1.4 \times 10^{-5}$, and get
down to $\Delta m^2\sim 0.08$ eV$^2$.

\section{Cosmological Constraints on Neutrinos}

We heard here~\cite{Olive} that the success of Big-Bang Nucleosynthesis
constrains the number of neutrino species:
\begin {equation}
N_{\nu} = 3.0 \pm 0.23 \pm 0.38^{+0.11}_{-0.57}
\label{bbnnnu}
\end{equation}
where the last uncertainty is that due to the baryon-to-entropy ratio
$\eta$.  We also heard~\cite{Olive} that there is no crisis for Big-Bang
Nucleosynthesis, but possibly for over-simplified theories of the chemical
evolution of the Galaxy~\cite{Hata}.  The range (\ref{bbnnnu})
corresponds
to an upper bound
\begin{equation}
N_{\nu} < 3.9 (90\% \hbox{c.l.})
\label{uppernnu}
\end{equation}
to be compared with the LEP determination $N_{\nu} = 2.989 \pm
0.012$~\cite{LEP}.

We also heard~\cite{Kainulainen} that Big-Bang Nucleosynthesis constrains
the mass of any metastable neutrino, assumed to be the $\tau$ neutrino:
\begin{equation}
m_{\nu_{\tau}} > 32 \hbox{MeV or} < 0.95 \hbox{MeV}
\label{limmnum}
\end{equation}
if it has a Dirac mass, and
\begin{equation}
m_{\nu {\tau}} > 25\hbox{MeV or} < 0.37 \hbox{MeV}
\label{limmnud}
\end{equation}
if it has a Majorana mass.  Both of the lower limits in
(\ref{limmnum},\ref{limmnud}) conflict with the upper bound
$m_{\nu_{\tau}} < 23$ MeV from ALEPH~\cite{Gregorio}, indicating that
$m_{\nu_{\tau}}$ must actually lie below $1$ MeV.  We also know that any
stable neutrino must weigh less than 
\begin{equation}
m_{\nu_{\tau}} = 92 h^2\hbox{eV}
\label{cosmomnu}
\end{equation}
where $h$ is the present expansion rate of the Universe in units of $100$
km/s/Mpc, and it is assumed that the present mass density of the Universe
is no larger than the critical density, $\Omega = 1$, as suggested by
theories of cosmological inflation. There are also cosmological 
constraints on unstable
neutrinos~\cite{Kainulainen}, which I will not go into here.

There has recently been discussion~\cite{Primack} about the observational
value of $h$ and its compatibility with the age of the Universe as
determined by astrophysicists.  We see from Fig.~7 that there is no
significant discrepancy~\cite{Primack}, provided $h$ is not much larger
than about $0.5$~\footnote{This is not incompatible with recent
astrophysical determinations of $h$, for example from time delays in
quasar lensing~\cite{macrolens}.}, corresponding to $m_{\nu_{\tau}}
\lappeq 23$ eV. On the
other hand, it may well be that most of the critical density is not in the
form of massive neutrinos.

\begin{figure}[H]
\hglue1cm
\epsfig{figure=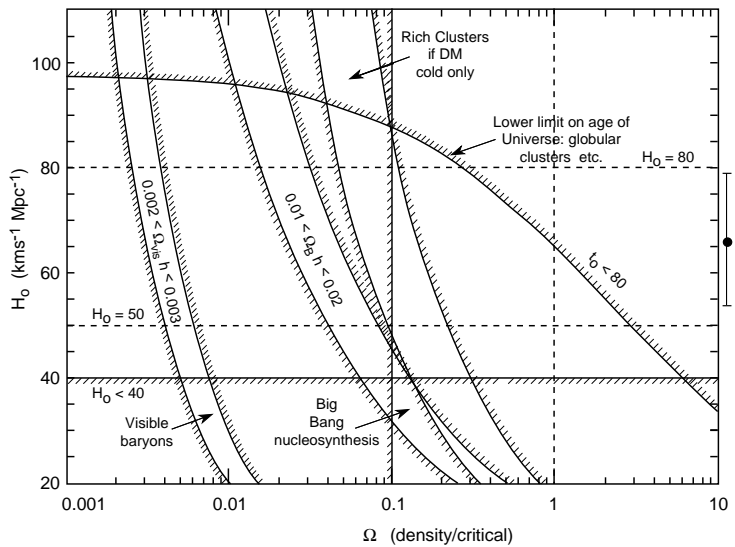,width=10cm}
\caption[]{The $(\Omega, H_0)$ plane exhibits no serious discrepancy between the
average measured value of $H_0$ (indicated by the vertical error bar), $\Omega = 1$ 
and an age for the Universe
of $10^{10}$ years. This plot also shows the estimates of the present
baryon density obtained from visible features in the Universe, from Big
Bang Nucleosynthesis and from rich clusters.
}
\end{figure}

The favoured theory of the formation of structures in the early Universe
is that they are due to the gravitational unstability of overdense regions
as they come within the horizon.  These overdense regions would have
originated from density perturbations created by quantum fluctuations
during the inflation.  However, it is generally thought that the growth of
such inflationary perturbations needs to be accelerated by matter that is
non-relativistic (cold) during the epoch of structure
formation~\cite{Primack}, as illustrated in Fig.~8.

It is generally thought that most of the present mass density must
be in the form of cold dark matter.  However, the detailed comparison of
data on microwave background fluctuations as first observed using the
COBE satellite~\cite{COBE}, other observational data on large-scale
astrophysical structures, and data on smaller scales, indicate that the
pure cold dark matter model requires modification.  One possibility is
that
there may be energy density in the vacuum (a cosmological constant),
another is that the spectrum of inflationary perturbations may be
scale-dependent (tilted), and a third is that there is an admixture of hot
dark matter~\cite{MDM}, i.e., matter that was relativistic during the
epoch of
structure formation:
\begin{equation}
\Omega_{Cold} \sim 0.7, \, \, \Omega_{Hot} \sim 0.2
\label{mixed}
\end{equation}
The only plausible candidate for the hot component of the dark matter is
one or more species of massive neutrino.

\begin{figure}[H]
\hglue2cm
\epsfig{figure=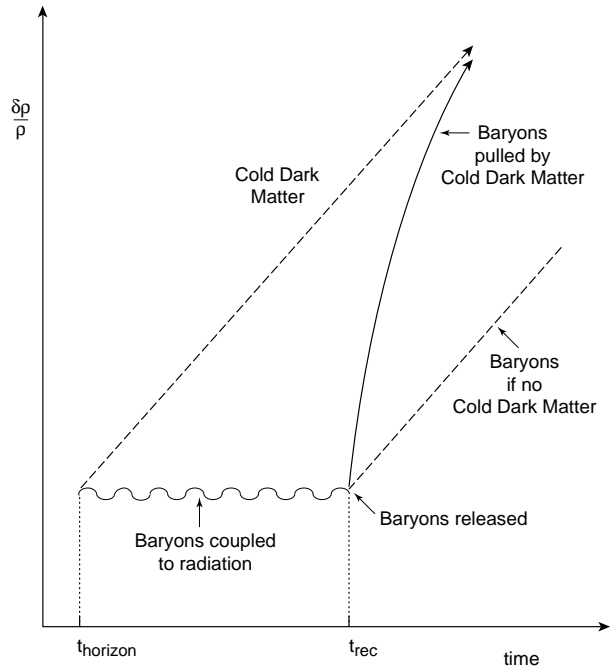,width=8cm}
\caption[]{A sketch indicating the r\^ole of Cold Dark Matter in magnifying
primordial
density perturbations via gravitational instability, while the baryons are
still coupled to radiation.}
\end{figure}

Simulations of structure formation indicate that any cosmological constant
is unlikely to be large enough to alleviate significantly the
(non-existent) age problem.  Moreover, peculiar motions
on the scale of
galactic clusters seem to be larger than expected in cold dark matter
models with tilt.  As summarized in Fig.~9, mixed dark matter models
(\ref{mixed}) seems to fit the observational data best~\cite{Primack}.
Although it has
been suggested that there may be more than one species of neutrino with
similar masses~\cite{PHKC}, the simplest hypothesis, 
which is consistent with the
seesaw mass matrix (\ref{seesaw}), is that one neutrino, most likely the
$\nu_{\tau}$, dominates the hot dark matter density.  In such a mixed dark
matter model, one would have
\begin{equation}
m_{\nu_{\tau}} \sim 5 \hbox{eV}
\label{mest}
\end{equation}
Moreover, such a mass (\ref{mest}) is quite consistent with the value of
$m_{\nu_{\mu}} \sim 2 \times 10^{-3}$ eV expected in the MSW
interpretation of the solar neutrino deficit, and the ratio $\sim
(m_t/m_c)^2$ expected in the simple seesaw model (\ref{seesaw}). If true,
the estimate (\ref{mest}) suggests an optimization of accelerator
searches
for neutrino oscillations down to $\Delta m^2 \sim 10$ eV$^2$.

\begin{figure}[H]
\hglue2cm
\epsfig{figure=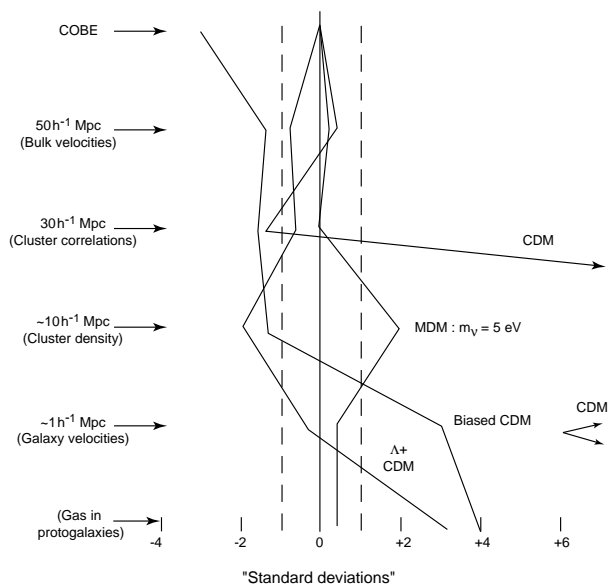,width=8cm}
\caption[]{A sketch indicating the relative successes of different models of
structure formation, as compared with different types of astrophysical and
cosmological data~\cite{Primack}.
}
\end{figure}
\section{Neutralinos}

We have already addressed the possibility that a massive neutrino might
constitute hot dark matter in the Universe.  Whether this is necessary is
still an open question: what seems much better established is the need for
a large amount of cold dark matter, at least on a cosmological scale if
inflation is to be accepted, and also on a galactic scale if models of
structure formation are to be believed. The question of immediate
experimental
interest is how much dark matter of what type may be present in the
galactic halo that surrounds us.  This is presumably not constituted of
neutrinos, because they would not cluster on such a small scale~\cite{ES}.
Recently a couple of observational programmes have seen microlensing
events interpreted as due to sub-solar-mass objects in the galactic
halo~\cite{Milsztayn}.  The best estimate of the MACHO
collaboration~\cite{MACHO} is
that about half our halo could be in this baryonic form, and perhaps even
all of it, whereas the EROS collaboration~\cite{EROS} seems to rule out
this
probability.  Even if $60\%$ or more of our galactic
halo is in the form of microlensing objects, one may anticipate
a local
density of cold dark matter particles of $10^{-25}$ gcm$^{-3}$ or
more~\cite{TGG}.

My favourite candidate for this cold dark matter is the lightest
supersymmetric particle, usually thought to be a neutralino, i.e., some
mixture of the spin-$1/2$ supersymmetric partners of the $Z^0$, $\gamma$
and Higgs~\cite{EHNOS}.  Fig.~10 shows the lower limit on the neutralino
mass established (modulo certain loopholes) by the ALEPH
collaboration~\cite{ALEPH} on the basis of unsuccessful sparticle searches
at LEP 1 and 1.5.  There is a good
likelihood~\cite{JKG} that the cosmological relic density of neutralinos
may lie in the range of interest to inflationary cosmologists: $0.1
<\Omega h^2 <0.3$~\cite{Berez}.  As also seen in Fig.~10, the ALEPH
limit~\cite{ALEPH} may be strengthened (and its loopholes removed) if one
postulates that the neutralino density falls within this
range~\cite{EFOS}~\footnote{For updates of this analysis to include the
preliminary results of higher-energy LEP 2 data, see~\cite{EFOS'}.}.

Both of these analyses~\cite{ALEPH,EFOS} are based on the assumption that
sparticle masses have certain universality properties which may not be
valid.  If these asumptions are relaxed, the lower bound on the mass of 
the lightest neutralino may not be altered qualitatively, but its
phenomenology may be significantly modified~\cite{Berez}, perhaps changing
from a mainly gaugino composition to a mainly higgsino composition as seen
in Fig.~11.  Such a change can also alter the prospects for neutralino
searches, as we discuss next.

Three strategies for such searches are favoured: neutralino annihilation
in the galactic halo which yields stable particles (${\bar p}, e^+,
\gamma, \nu$) in the cosmic rays~\cite{halo}, 
which will be explored by the AMS satellite, annihilation after capture
within the Sun or Earth, which may yield high-energy neutrinos detectable
in underground detectors (either directly or via the $\mu$s generated by
$\nu$ interactions in rock)~\cite{trap}, and elastic scattering on a
nuclear target in the laboratory~\cite{GW}.
\vfill\eject

\begin{figure}[H]
\hglue3cm
\epsfig{figure=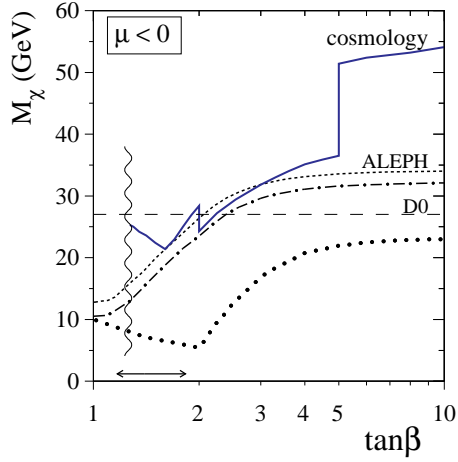,width=6cm}
\caption[]{The ALEPH lower limit~\cite{ALEPH} on the lightest neutralino mass
$m_{\chi}$ (short-dashed line) as a
function of the ratio of Higgs vacuum expectation values tan$\beta$,
which has an important loophole indicated by the double arrow, is compared
with the absolute lower limits obtained from a combination with other
$e^+ e^-$ experiments (dotted line), with that inferred from the D0
experiment (long-dashed line), and those obtained from combining
phenomenological and cosmological considerations, both with (solid line)
and without (dot-dashed line) the supplementary theoretical assumption
of dynamical electroweak symmetry breaking~\cite{EFOS}.
}
\end{figure}

\begin{figure}[H]
\hglue1.5cm
\epsfig{figure=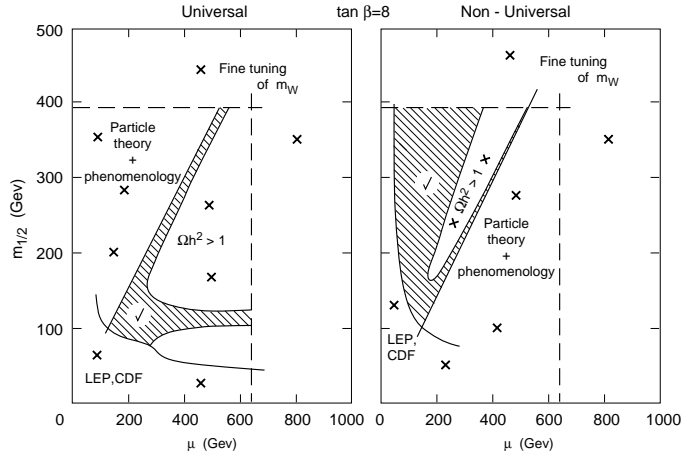,width=9cm}
\caption[]{The preferred composition of the lightest neutralino may change from
being mainly a gaugino to mainly a higgsino, if the normal assumptions of
universality are relaxed~\cite{Berez}.}
\end{figure}

In the second case, which is of particular interest to neutrino
physicists,
annihilation follows the passage of relic neutralinos through the Sun or
Earth accompanied by scattering and the loss of recoil energy, which
is what causes the
the neutralinos to become trapped~\cite{trap}.  Among the neutralino
annihilation products
will be some energetic neutrinos from $\tau$ or heavy-quark decays, which
can escape from the core of the Sun or Earth and be detected in
underground experiments, either directly or indirectly as mentioned above.
Underground $\mu$ search experiments such as Baksan are now imposing
significant constraints on models~\cite{Berez,Bergstrom}, as seen in
Fig.~12.  This type of search is promising for the future generation of
experiments that includes Baikal, Nestor and Amanda~\cite{AMANDA}.  One
point that needs
to be watched is the possibility that also these high-energy neutrinos
oscillate~\cite{Masood}. In particular, there could be a significant
suppression of the solar $\nu_{\mu}$ flux and a corresponding enhancement
of the $\nu_e$ flux if the large-angle MSW solution to the low-energy
solar-neutrino deficit is correct.  This implies the need for some caution
in interpreting upper limits on the flux of muons generated by high-energy
neutrinos from the Sun.

\begin{figure}[H]
\hglue2cm
\epsfig{figure=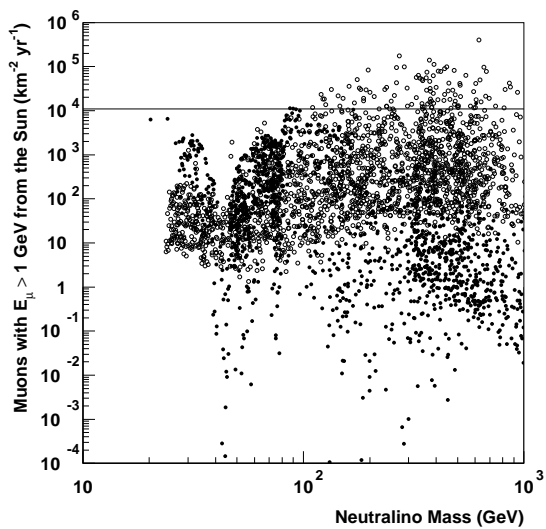,width=8cm}
\caption[]{Upper limits on energetic neutrino emission from the Sun due to
neutralino annihilations are already beginning to exclude certain
models~\cite{Bergstrom}.
}
\end{figure}

The direct detection of the elastic scattering of neutralinos on nuclei in
the laboratory~\cite{GW} is also a promising technique.  There are
important contributions to the scattering matrix elements from both
spin-dependent and spin-independent forces.  The former are determined by
axial-current matrix elements related~\cite{EF} to that appearing in
neutron $\beta$ decay~\footnote{It should be emphasized that the naive
quark model is not a good guide to the magnitudes of these matrix
elements, in particular since there is a significant contribution from the
${\bar s} \gamma_{\mu} \gamma_5 s$ current~\cite{EK}}, whereas the latter
are related to the different quark contributions to the nucleon mass.
These spin-independent forces are coherent for heavy nuclei, and likely to
dominate for favoured detector materials such as Ge~\cite{JKG}.  Recent
upper limits on the spin-dependent and spin-independent elastic
dark-matter scattering rates were shown at this meeting~\cite{scatter}.
Eventually, such
experiments should improve significantly in sensitivity, and may be able
to compete with, and complement, accelerator searches for supersymmetric
particles~\cite{EF}.

\section{The CERN Experimental Programme}

The large accelerator laboratories such as Fermilab and CERN have not been
very much in evidence at this meeting, reflecting the long-term trend of
neutrino physics towards non-accelerator experiments.  Nevertheless,
accelerator
laboratories do have important roles to play, and I would like to mention
briefly some of the neutrino and neutrino-related activities at my home
laboratory, CERN.

As you know, the present CERN short-baseline neutrino oscillation
experiments,  CHORUS~\cite{CHORUS} and NOMAD~\cite{NOMAD}, will
continue taking
data until the end of 1997.  Two other neutrino-related experiments have
also been taking data during 1996.  One is SPY~\cite{SPY}, which has
measured charged-particle yields in $p$-Be collisions, with a view to
the better calibration of accelerator $\nu$ beams and perhaps better
predictions for atmospheric $\nu$ fluxes. The other is NA55~\cite{NA55},
which has
been measuring neutron production in $\mu$ collisions, with a view to
understanding better this possible background for atmospheric $\nu$
experiments~\cite{bgrd}.

Another activity at CERN is COSMOLEP, which seeks to
use the $\mu$ detectors of LEP experiments to look at extended air
showers~\cite{Cosmolep}.  The use of L3 to measure the atmospheric
$\mu$ flux has also been approved recently~\cite{cosmoL3}.  This could
reduce
significantly the current uncertainties in the flux, enabling one to
determine whether Kamiokande and IMB have been seeing too few $\nu_{\mu}$
and/or too many $\nu_e$~\cite{Gaisser}.

There are also extensive
discussions taking place at CERN about the possibility of one or more
future $\nu$ experiments following CHORUS and NOMAD.  One of these
possibilities is an idea to use a $\nu$ beam at the PS accelerator
to probe the LSND claim~\cite{PS}.  Another possibility is a 
short-baseline
experiment at the CERN SPS, as a follow-up to CHORUS and
NOMAD~\cite{short}.
Talk of a medium-baseline experiment using a detector in the Jura
mountains, about $20$km away from CERN, has also been revived
recently~\cite{medium}.
Another attractive option would be a long-baseline
experiment~\cite{long}
sending a beam to the Gran Sasso laboratory, $730$ km away, which might be
accompanied by a nearby detector near or under the Geneva airport.

In my view, one of the most interesting options on this list is the
short-baseline follow-up to CHORUS and NOMAD.  The COSMOS
experiment~\cite{COSMOS}
planned for Fermilab is very promising, but it is a long time in the
future, and, as I emphasized earlier, one should always require
confirmation in the neutrino business.  On the other hand, there are
already two entries in the long-baseline race, LBLE~\cite{Suzuki} and
MINOS~\cite{MINOS}. Moreover,
the motivations for neutrino oscillations detectable in 
a suitable short-baseline experiment,
namely hot dark matter, the seesaw mechanism and the solar neutrino
deficit, remain just as strong as ever. The only change, perhaps, is a
trend towards a slightly lower neutrino mass for the cosmological hot dark
matter. As already mentioned, this might suggest a somewhat lower neutrino
energy and/or longer
baseline for any follow-up experiment, which is motivated whether or not
CHORUS and NOMAD find anything in their present data sets.

\section{Conclusion}

To conclude this talk, I present a possible chronology of future neutrino
experiments and others of potential interest to this community.  An
exciting era is opening up, with major new experiments such as
Superkamiokande and LEP 2 starting to take data.  Some of the major issues
in particle physics have a chance of being resolved by the time of
Neutrino 98, with the opening up of new domains of exploration for the
Higgs boson, supersymmetric particles and neutrino masses.  Let us all
cross all amenable body parts, and hope for progress by the next meeting
in this series!
\vfill\eject

\begin{table}[H]
\begin{center}
Possible Chronology of Future Neutrino and Related Experiments\\
\bigskip
\begin{tabular}{|c|c|c|c|} \hline
Year  & Accelerator  & Other       & Non-accelerator \\
~~~~  & Neutrino     & Accelerator & Experiments     \\
~~~~  & Experiments  & Experiments & ~~~~            \\
\hline\hline
1996  & CHORUS,     & LEP 2, SPY,     & AMANDA, Baikal,  \\
~~~~  & NOMAD       & COSMOLEP,  & SKK, Chooz,      \\
~~~~  & ~~~~~       & NA55            & Homestake Iodine \\ \hline
1997  & LSND $\nu$  & CosmoL3         & SNO,             \\
~~~~  & ~~~~        & ~~~~~~          & Palo Verde       \\ \hline
1998  & KARMEN upgrade & ~~~~~~          & AMS,             \\
~~~~  & ~~~~        & ~~~~~~          & GNO?             \\ \hline
1999  & LBLE(12 GeV) & B factories,   & BOREXINO,        \\
~~~~  & ~~~~        & HERA-B          & ICARUS 600t      \\ \hline
2000  & ALADINO/NOE/ & ~~~~~          & $0 \nu 2 \beta \rightarrow 0.2$
eV? \\ 
~~~~  & TENOR/HELLAZ/ & ~~~~          & ~~~~~            \\ 
~~~~  & full ICARUS?  & ~~~~~~          & ~~~~~            \\ \hline
2001  & COSMOS,     & ~~~~~~          & MAP?             \\
~~~~  & MINOS       & ~~~~~~          & ~~~~~            \\ \hline
2002  & ~~~~~       & ~~~~~~          & ~~~~~            \\ 
~~~~  & ~~~~~       & ~~~~~~          & ~~~~~            \\ \hline
2003  & LBLE(50 GeV) & ~~~~~          & COBRAS/SAMBA?    \\ 
~~~~  & ~~~~~       & ~~~~~~          & ~~~~~            \\ \hline
2004  & ~~~~~       & ~~~~~~          & BAND?            \\ 
~~~~  & ~~~~~       & ~~~~~~          & ~~~~~            \\ \hline
2005  & ~~~~~       & LHC             & $0 \nu 2 \beta \rightarrow 0.1$
eV? \\
~~~~  & ~~~~~       & ~~~~~           & ~~~~~            \\ \hline
\hline
\end{tabular}
\end{center}
\end{table}



\begin{thebibliography}{99}


\bibitem{LEP} A.~Blondel, Plenary talk at the {\it International
                Conference
                on High Energy Physics}, Warsaw, 1996, reporting the
                analysis 
		of the LEP Electroweak Working Group and the SLD Heavy
		Flavor Group, CERN Report No.\ LEPEWWG/96-02, available at
		the URL: http://www.cern.ch/LEPEWWG.

\bibitem{PDG}  Particle Data Group, R.~M.~Barnett et al.,
		\PR {\bf D54} (1996) 1.

\bibitem{discrep} P. Vogel, V. Sandberg, talks at this meeting.

\bibitem{nup} J. Ellis and M. Karliner, \PL {\bf B213} (1988) 73;\\
D.B. Kaplan and A. Manohar, \NP {\bf B310} (1988) 527.

\bibitem{CCFR} R. Bernstein, talk at this meeting;\\
A.L. Kataev, A.V. Kotikov, G. Parente and A.V. Sidorov, \PL {\bf B388}
(1996) 179.

\bibitem{CDFD0} M. Demarteau, Fermilab preprint Conf-96/354, 
hep-ex/9611019, and references therein.

\bibitem{LEP2} ALEPH, DELPHI, L3 and OPAL collaborations,
presentations at open session of the CERN LEP Experiments
Committee, Oct. 8th, 1996;\\
See, for example, OPAL collaboration, K. Ackerstaff et al., 
CERN Preprint PPE/96-141.

\bibitem{MH} J. Ellis, G.L. Fogli and E. Lisi, CERN preprint
TH/96-216
(1996), hep-ph/9608329; \\ see also W. de Boer, A. Dabelstein, W. Hollik,
W. M\"osle and U. Schwickerath, hep-ph/9607286;\\
S. Dittmaier and D. Schildknecht, hep-ph/9609488.

\bibitem{GLS} A.L. Kataev and A.V. Sidorov, \PL {\bf B331}
(1994) 179;\\ 
CCFR-NUTEV collaboration, D. Harris et al., hep-ex/9506010;\\
see also L.S. Barabash et al., hep-ex/9611012.

\bibitem{Schmelling} M. Schmelling, rapporteur talk at International
Conference on High-Energy Physics, Warsaw 1996.

\bibitem{tau} M. Girone and M. Neubert, \PRL {\bf 76} (1996) 3061, and
references therein.

\bibitem{EGKS} J. Ellis, E. Gardi, M. Karliner and M. Samuel, \PL {\bf
B366} (1996) 268 and \PR {\bf D54} (1996) 6986.

\bibitem{seesaw}  T. Yanagida, {\it Proc. Workshop on the Unified Theory and the
Baryon Number in the Universe} (KEK, Japan, 1979);\\ R. Slansky, Talk at {\it
Sanibel Symposium}, Caltech Preprint CALT-68-709 (1979).

\bibitem{Bonn} J. Bonn, talk at this meeting. The possibility of a
crumpled configuration of the (supposedly) monolayer source, leading to
multiple scattering, is a concern in this experiment.

\bibitem{Lobashev} V.M. Lobashev, talk at this meeting. I
understand from A. Kusenko that a re-evaluation of multiple scattering in
the source of this experiment may lead to a revised upper limit on the
possible high-mass branch of about $0.5 \%$.

\bibitem{Swift} A. Swift, talk at this meeting.

\bibitem{KKGH} H.V. Klapdor-Kleingrothaus, talk at this meeting.

\bibitem{betabeta} F.T. Avignone, A. Barabash, H. Ejiri, J. Farine and E.
Fiorini, talks at this meeting.

\bibitem{Home} K. Lande, talk at this meeting;

\bibitem{Suzuki} Y. Suzuki, talk at this meeting:\\
Kamiokande collaboration, Y. Fukuda et al., preprint ICRR-372-96-23
(1996);\\
see also J.N. Bahcall, P.I. Krastev and E. Lisi, Princeton preprint
IASSNS-AST 96/43.

\bibitem{Gavrin} V.N. Gavrin, talk at this meeting;\\
SAGE Collaboration, D.N. Abdurashitov et al., \PL {\bf B328} (1994) 234.

\bibitem{Kirsten} T. Kirsten, talk at this meeting;\\
GALLEX Collaboration, W. Hampel et al., Saclay preprint DAPNIA-SPP-96-10
(1996), and references therein.

\bibitem{IMB} IMB Collaboration, R. Becker-Szendy et al., \PR {\bf D46}
(1992) 3720.

\bibitem{Soudan} E.A. Peterson, talk at this meeting; \\
Soudan I Collaboration, W.W.M. Allison et al., hep-ex/9611007.

\bibitem{others} NUSEX Collaboration, M. Aglietta et al., {\it Europhys.
Lett.} {\bf 8}
(1989) 611;\\
Fr\'ejus Collaboration, C. Berger et al., \PL {\bf B227} (1989) 489 and
{\bf B245} (1990) 305.

\bibitem{Caldwell} D.O. Caldwell, talk at this meeting;\\
LSND Collaboration, C. Athanassopoulos et al., nucl-ex/9605001 and
nucl-ex/9605003.

\bibitem{Bahcall} J. Bahcall, talk at this meeting;\\
J. Bahcall and M.H. Pinsonneault, hep-ph/9610542.

\bibitem{Gaisser} T. Gaisser, talk at this meeting, hep-ph/9611301;\\
T. Gaisser, M. Honda, K. Kasakara, H. Lee, S. Midorikawa, V. Naumov and T.
Stanler, hep-ph/9608253.

\bibitem{Fogli} G.L. Fogli and E. Lisi, \PR {\bf D52} (1995) 2775.

\bibitem{KARMEN} J. Kleinfeller, talk at this meeting;\\
KARMEN Collaboration, B. Bodmann et al., \PL {\bf B332} (1994) 251 and
{\bf B348} (1995) 19.

\bibitem{CCFRosc} CCFR Collaboration, A. Romosan et al., Columbia
preprint NEVIS-1529, hep-ex/9611013.

\bibitem{Smirnov} A. Smirnov, talk at this meeting

\bibitem{Fior} See, e.g., N.  Hata, S.A. Bludman and P. Langacker, 
\PR {\bf D49} (1994) 3622;\\
V. Castellani, S. Degl'Innocenti and G.
Fiorentini, \ASAS {\bf 271} (1993) 601.

\bibitem{Conforto} G. Conforto, talk at this meeting.

\bibitem{Dar} A. Dar, talk at this meeting, astro-ph/9611014.

\bibitem{Betal} J.N. Bahcall, M. Pinsonneault, S. Basu and J.
Christensen-Dalsgaard, astro-ph/9610250.

\bibitem{helio} J. Christensen-Dalsgaard, \NP {\bf B} (Procl.Suppl.) {\bf
48} (1996) 325.

\bibitem{Petcov} S. Petcov, talk at this meeting.

\bibitem{MSW} L. Wolfenstein, \PR {\bf D17} (1978) 2369;\\
S.P. Mikheyev and A.Yu. Smirnov, \YF {\bf 42} (1985) 1441.

\bibitem{Krastev} P. Krastev, private communication.

\bibitem{Burgess} C. Burgess, talk at this meeting;\\
C.P. Burgess and D. Michaud, hep-ph/9606295 and hep-ph/9611368.

\bibitem{Rossi} A. Rossi, talk at this meeting;\\
H. Nunokawa, A. Rossi, V. Semikoz and J.W.F. Valle, \NP {\bf B472}
(1996) 495 and hep-ph/9610526.

\bibitem{Rowley} K. Rowley, public communication.

\bibitem{SNO} A. McDonald, talk at this meeting;\\
SNO Collaboration, G.T. Evan et al., SNO-87-12 (1987);\\
see also J.N. Bahcall and E. Lisi, Princeton Preprint IASSNS-AST 96/33
(1996).

\bibitem{BOREXINO} E. Bellotti, talk at this meeting;\\
BOREXINO Collaboration, C. Arpesella et al., BOREXINO proposal
(University of Milano, Milano, 1992);\\
see also J.N. Bahcall and P.I.
Krastev, astro-ph/9607013.

\bibitem{ICARUS} ICARUS Collaboration, P. Cennini et al., ICARUS proposal
(1993).

\bibitem{HELLAZ} HELLAZ Coolaboration, G. Laurenti et al., {\it Proc.
Fifth Int. Workshop on Neutrino Telescopes}, Venice 1993, ed. M.
Baldo-Ceolin (University of Padova, Padova, 1994) 161.

\bibitem{SPY} SPY Collaboration, http://www.cern.ch/NA56/ .

\bibitem{cosmoL3} P. Le Coultre, on behalf of the L3 collaboration, 
private communication.

\bibitem{MINOS} S. Wojcicki, talk at this meeting;\\
MINOS Collaboration, E. Ables et al., Fermilab proposal P-875 (1995).

\bibitem{long} NOE Collaboration, M. Ambrosio et al., {\it Nucl. Inst.
Meth.} {\bf A363} (1995) 604;\\
T. Ypsilantis, {\it Nucl. Inst. Meth.} {\bf A371} (1996) 330;\\
see also~\cite{ICARUS}.

\bibitem{Chooz} C. Bemporad, talk at this meeting;\\
see also R.I. Steinberg, hep-ph/9306282.

\bibitem{Palo}G. Gratta, talk at this meeting;\\
Palo Verde Collaboration, F. B\"ohm et al., Stanford preprint HEP-96-04
(1996).

\bibitem{Hill} J.E. Hill, \PRL {\bf 75} (1995) 2654.

\bibitem{PSI} M. Daum, talk at this meeting.

\bibitem{Primack} J. Primack, talk at this meeting;\\
see also J. Primack, hep-ph/9610321;\\
J. Primack and A. Klypin, astro-ph/9607061.

\bibitem{CHORUS} K. Niwa, talk at this meeting;\\ 
CHORUS Collaboration, http://choruswww.cern.ch/ .

\bibitem{NOMAD} J. Dumarchez, talk at this meeting;\\ 
NOMAD Collaboration, http://nomadinfo.cern.ch/ .

\bibitem{COSMOS} R.A. Sidwell, talk at this meeting;\\
COSMOS Collaboration, P803 proposal to Fermilab (1993).

\bibitem{Olive} K. Olive, talk at this meeting;\\
see also: K. Olive, astro-ph/9609071;\\
K. Olive and D. Thomas, hep-ph/9610319;\\
for the latest words on the primordial lithium abundance, see
P. Bonifacio and P. Monaro, astro-ph/9611043;\\
and on the controversial primordial deuterium abundance,
see A.~Songaila, E.J.~Wampler and L.L.~Cowie, astro-ph/9611143.

\bibitem{Hata} N. Hata et al., \PRL {\bf 77} (1995) 3977;\\
G. Steigman, astro-ph/9608084.

\bibitem{Kainulainen} K. Kainulainen, talk at this meeting;

\bibitem{Gregorio} A. Gregorio, talk at this meeting;\\ 
ALEPH Collaboration, D. Buskulic et al., \PL {\bf B349} (1995) 585.

\bibitem{macrolens} T. Kundi\'c et al., astro-ph/9610162.

\bibitem{COBE} C.L. Bennett et al., astro-ph/9601067;\\
future satellite projects are MAP and COBRAS/SAMBA.

\bibitem{MDM} R. Schaefer and Q. Shafi, {\it Nature} {\bf 359} (1992)
119;\\
M. Davis, F.J. Summers and D. Schlegel, {\it Nature} {\bf 359} (1992)
393;\\
A.N. Taylor and M. Rowan-Robinson, {\it Nature} {\bf 359} (1992) 396.

\bibitem{PHKC} J. Primack, J. Holtzman, A. Klypin and D.O. Caldwell, \PRL
{\bf 74} (1995) 2160;\\
see also S. Ghigna et al., astro-ph/9611103.

\bibitem{ES} J. Ellis and P. Sikivie, \PL {\bf B321} (1994) 390.

\bibitem{Milsztayn} A. Milsztayn, talk at this meeting;\\
see also: F. Adams and G. Laughlin, astro-ph/9602006;\\
G. Chabrier, L. Segretain and D.M\'era, astro-ph/9606083;\\ 
A.~Gould, astro-ph/9611185.

\bibitem{MACHO} MACHO collaboration, C. Alcock et al., astro-ph/9606165.

\bibitem{EROS} EROS collaboration, R. Ansari et al., \ASAS {\bf 299} (1995)
L21.

\bibitem{TGG} M.S. Turner, E.J. Gates and G. Gyuk, astro-ph/9601168.

\bibitem{EHNOS} J. Ellis, J.S. Hagelin, D.V. Nanopoulos, K.A. Olive
and M. Srednicki, \NP {\bf B238} (1984) 453.

\bibitem{ALEPH} ALEPH collaboration, D. Buskulic et al., CERN preprint
PPE/96-83 (1996).

\bibitem{EFOS} J. Ellis, T. Falk, K.A. Olive and M. Schmitt, \PL
{\bf B388} (1996) 97.

\bibitem{JKG} G. Jungman, M. Kamionkowski and K. Griest, {\it Phys. Rep.}
{\bf 267} (1996) 195.

\bibitem{Berez} V. Berezinskii, talk at this meeting, astro-ph/9610263;\\
see also A. Bottino, N. Fornengo, G. Mignola, M. Olechowski and S.
Scopel, astro-ph/9611030.

\bibitem{EFOS'} J. Ellis, T. Falk, K.A. Olive and M. Schmitt, CERN
preprint TH/96-284, hep-ph/9610410.

\bibitem{halo} See, for example, J. Silk and M. Srednicki, \PRL {\bf 53}
(1984) 624;\\
J. Ellis et al., \PL {\bf B214} (1988) 403.

\bibitem{trap} J. Silk, K.A. Olive and M. Srednicki, \PRL {\bf 55} 1985)
259.

\bibitem{GW} M. Goodman and E. Witten, \PR {\bf D31} (1985) 3059.

\bibitem{Bergstrom} L. Bergstrom, talk at this meeting;\\
L. Bergstr\"om, J. Edsj\"o and P. Gondolo, hep-ph/9607237.

\bibitem{AMANDA} F. Halzen, for the AMANDA collaboration, hep-ex/9611014.

\bibitem{Masood} J. Ellis, R.A. Flores and S. Masood, \PL {\bf B294}
(1992) 229.

\bibitem{EF} J. Ellis and R.A. Flores, \PL {\bf B263} (1991) 259 and {\bf
B300} (1993) 175; \NP {\bf B307} (1988) 375 and {\bf B400} (1993) 25.

\bibitem{EK} J. Ellis and M. Karliner, Lectures at the {\it Int. School of
Nucleon Spin Structure}, Erice 1995, CERN preprint TH/95-334,
hep-ph/9601280.

\bibitem{scatter} F.T. Avignone, talk at this meeting;\\
J.J. Quenby et al., Imperial College, London, Astrophysics Group preprint 5
(1996).

\bibitem{NA55} NA55 Collaboration, F. B\"ohm et al., CERN SPSLC/95-62,
Proposal P293 (1995).

\bibitem{bgrd} O.G. Ryazhskaya, \NC {\bf C18} (1995) 77;\\
{\it J.E.T.P. Lett.} {\bf 61} (1995) 237;\\
see, however, Kamiokande Collaboration, Y. Fukuda et al., Preprint
ICRR-373-96-24 (1996).

\bibitem{Cosmolep} COSMOLEP Collaboration, http://alephwww.cern.ch/
COSMOLEP/ .

\bibitem{PS} P. Zucchelli, in preparation.

\bibitem{short} ALADINO Collaboration, M. Bonesini et al., CERN
SPSLC/95-37, Letter of Intent I205 (1995);\\
J.-J. Gomez-Cardenas et al., {\it Nucl. Inst. Meth.} {\bf A378} (1996)
196;\\
TENOR Collaboration, A. Ereditato et al., CERN preprint PPE/96-106 (1996).

\bibitem{medium} See T. Ypsilantis, second paper in~\cite{long}.

\end{thebibliography}
\end{document}